\documentclass[conference]{IEEEtran}
\IEEEoverridecommandlockouts
\usepackage{cite}
\usepackage{amsmath,amssymb,amsfonts}
\usepackage{graphicx}
\usepackage{textcomp}
\usepackage{xcolor}
\usepackage{algorithm}
\usepackage{algpseudocode}
\usepackage{ulem}

\def\BibTeX{{\rm B\kern-.05em{\sc i\kern-.025em b}\kern-.08em
    T\kern-.1667em\lower.7ex\hbox{E}\kern-.125emX}}
\begin{document}
\renewcommand{\slash}{/\penalty\exhyphenpenalty\hspace{0pt}}
\newenvironment{ldescription}[1]
  {\begin{list}{}%
   {\renewcommand\makelabel[1]{##1\hfill}%
   \settowidth\labelwidth{\makelabel{#1}}%
   \setlength\leftmargin{\labelwidth}
   \addtolength\leftmargin{\labelsep}}}
  {\end{list}}

\title{Solving scalability issues in calculating PV hosting capacity in low voltage distribution networks
\thanks{[This work has been supported in part by the The Recovery and Resilience Facility (RRF) under NPOO.C1.6.R1-I2.01 project "Development of CharGo! products in value chain for tourism".}
}

\author{\IEEEauthorblockN{Tomislav Antić}
\IEEEauthorblockA{\textit{University of Zagreb}\\
\textit{Faculty of Electrical Engineering}\\\textit{and Computing} \\
Zagreb, Croatia \\
tomislav.antic@fer.hr}

\and

\IEEEauthorblockN{Alireza  Nouri, Andrew Keane}
\IEEEauthorblockA{\textit{University College Dublin}\\
\textit{UCD Energy Institute} \\
Dublin, Ireland \\
{alireza.nouri, andrew.keane}@ucd.ie}

\and

\IEEEauthorblockN{Tomislav Capuder}
\IEEEauthorblockA{\textit{University of Zagreb}\\
\textit{Faculty of Electrical Engineering}\\\textit{and Computing} \\
Zagreb, Croatia \\
tomislav.capuder@fer.hr}
}

\maketitle

\begin{abstract}
The share of end-users with installed rooftop photovoltaic (PV) systems is continuously growing. Since most end-users are located at the low voltage (LV) level and due to technical limitations of LV networks, it is necessary to calculate PV hosting capacity. Most approaches in calculating a network's hosting capacity are based on three-phase optimal power flow (OPF) formulations. Linearized and relaxed three-phase OPF formulations respectively lose their accuracy and exactness when applied to solve the hosting capacity problem, and only non-linear programming (NLP) models guarantee the exact solution. Compared to linearized or relaxed models, NLP models require a higher computational time for finding an optimal solution. The binary variables uplift the problem to mixed-integer (MI)NLP and increase the computational burden. To resolve the scalability issues in calculating the hosting capacity of single-phase connected PVs, we propose a method that does not entail binary variables but still ensures that PVs are not connected to more than one phase at a time. Due to a risk of a sub-optimal solution, the proposed approach is compared to the results obtained by the MINLP formulation. The comparison includes values of the solution time and technical quantities such as network losses, voltage deviations, and voltage unbalance factor.
\end{abstract}


\section{Introduction}
\subsection{Motivation and Literature Review}
Installation of photovoltaic (PV) units in low voltage (LV) distribution networks ensures consumption of locally generated electricity and helps decrease end-users electricity costs \cite{9354638}. At the same time, the installation of PVs is often uncoordinated, negatively affecting technical conditions and complicating the planning and operation of distribution networks \cite{en14010117}. It is necessary to determine the maximum power of PVs that can be installed in a network to prevent uncoordinated integration, i.e., PV hosting capacity that will ensure the increase in the share of PVs without violating any network's technical constraints \cite{en13184756}.

Most of the methods for calculating PV hosting capacity are based on the power flow simulations but other approaches can also successfully resolve the given problem \cite{KOIRALA2022111899, RAJABI2022112365}. When observing LV networks, power flow formulations become much more complex due to the network structure, mutual effects of phases, relatively high R/X ratio, and additional constraints such as voltage unbalance. Therefore, iterative approaches such as the one presented by Joshi and Gokaraju \cite{8586196} may take time to find the solution. Due to numerous uncertainties in LV distribution networks, deterministic methods, as one described by Hashemi et al. \cite{7593314} are being replaced by stochastic approaches \cite{MULENGA2021106928}. Stochasticity and uncertanties in an LV network further increase the complexity of a given problem. Some papers put forward scenario reduction techniques to mitigate the problem \cite{KOIRALA2022108535}. There is a lack of PV hosting capacity methods that are based on non-linear (NLP) optimal power flow (OPF) models, mostly due to the complexity of the formulation and the larger computational time needed to find the optimal solution. Therefore, other formulations, including approximation or linearization, are used for solving OPF problems \cite{optimization_models}. Despite the benefits of using other formulations, NLP models are the only ones that ensure the exact solution without any loss of accuracy. Linearized formulations lose accuracy when the operating point varies by a great deal and hence, do not suit the intended application. For relaxed formulations to work, some conditions should hold. Specifically, the objective function of the hosting capacity problem poses challenges to the application of relaxed formulations. In an unbalanced system, the mutual effects of phases, upper bound voltage limits (which is specifically important in the hosting capacity problem), the limit on unbalance factor, and modelling the discrete control devices, e.g., voltage regulators, may put the relaxation far from being exact. To bridge the identified gap, hosting capacity in this paper is evaluated using an exact NLP formulation of the OPF model.

Another identified gap in the papers that observe a single-phase connection of PVs is the lack of constraints that ensure such a connection. The methodology proposed by Ceylan et al. \cite{8260243} is based on randomization and an iterative process that does not guarantee an optimal phase selection and optimal solution. A single-phase connection of PVs is also observed by Mulenga et al. \cite{MULENGA2021106928}, and uncertainties related to parameters, including the connection phase, are modelled using a stochastic approach. We propose an approach that introduces binary variables to determine the optimal connection phase. However, this uplifts the problem to mixed integer (MI)NLP, increasing complexity of the problem. Therefore, four other formulations that are based on randomization and predetermining the connection phase and do not include binary variables are presented in our work. Solutions obtained by different approaches for the same case study are found, and the proposed approach is validated by comparing various variables and quantities between the solution of the proposed approach and an exact MINLP approach.

\subsection{Contributions}
To summarize, the main contributions of the paper are:
\begin{itemize}
    \item We use an exact NLP formulation of the three-phase OPF model in calculating single-phase PV hosting capacity.
    \item The problem is uplifted to the MINLP formulation in order to ensure the accurate solution that is to be used as the benchmark.
    \item Finally, selection of the optimal connection phase using binary variables is replaced with randomization process. The solution is compared to the one found using the accurate MINLP formulation in terms of the computational time and accuracy.
\end{itemize}

The rest of the paper is organized as follows: the OPF mathematical model and methodology for determining a PV system connection phase is presented in Section \ref{sec:methodology}, case studies are described in Section \ref{sec:cs}, results are shown in Section \ref{sec:results}, and final conclusions are given in Section \ref{sec:conclusion}.

\section{Methodology} \label{sec:methodology}
The problem of calculating the PV hosting capacity is modelled with a non-convex, non-linear three-phase OPF formulation, implemented in the tool presented in \cite{ppopf}. The OPF formulation used for calculating the hosting capacity of single-phase connected PVs is modelled as the current-voltage formulation.

\subsection{Optimal Power Flow Model}
Real and imaginary part of voltage drop of all phases $p$ for a branch $l$ connecting nodes $i$ and $j$ is calculated using (\ref{eq:voltage_drop_re}) and (\ref{eq:voltage_drop_im}):

\begin{equation}
    U_{j,p,t}^{re} = U_{i,p,t}^{re} - \sum_{q \in \{a,b,c\}}R_{l,pq}\cdot I_{l,ij,q,t}^{re} + \sum_{q \in \{a,b,c\}}X_{l,pq}\cdot I_{l,ij,q,t}^{im}   
\label{eq:voltage_drop_re}
\end{equation}

\begin{equation}
    U_{j,p,t}^{im} = U_{i,p,t}^{im} - \sum_{q \in \{a,b,c\}}R_{l,pq}\cdot I_{l,ij,q,t}^{im} - \sum_{q \in \{a,b,c\}}X_{l,pq}\cdot I_{l,ij,q,t}^{re}  
\label{eq:voltage_drop_im}
\end{equation}

Active and reactive power flow in branch $l$, connecting nodes $i$ to $j$ are constrained with (\ref{eq:line_act_power_calc_phase}) and (\ref{eq:line_react_power_calc_phase}).
\begin{equation}
    P_{l,ij,p,t} = U_{i,p,t}^{re}\cdot I_{l,ij,p,t}^{re} + U_{i,p,t}^{im}\cdot I_{l,ij,p,t}^{im} 
    \label{eq:line_act_power_calc_phase}
\end{equation}
\begin{equation}
    Q_{l,ij,p,t} = U_{i,p,t}^{im}\cdot I_{l,ij,p,t}^{re} - U_{i,p,t}^{re}\cdot I_{l,ij,p,t}^{im} 
    \label{eq:line_react_power_calc_phase}
\end{equation}

Real and imaginary part of currents of loads and PVs are calculated with (\ref{eq:p_element}) and (\ref{eq:q_element}). These expressions are valid for all nodes $n$ in an LV network.
\begin{equation}
    P_{n,p,t}^{load/PV} = U_{n},p,t^{re} \cdot (I_{n,p,t}^{load/PV})^{re} + U_{n,p,t}^{im} \cdot (I_{n,p,t}^{load/PV})^{im}
\label{eq:p_element}
\end{equation}
\begin{equation}
   Q_{n,p,t}^{load/PV} = U_{n,p,t}^{im} \cdot (I_{n,p,t}^{load/PV})^{re} - U_{n,p,t}^{re} \cdot (I_{n,p,t}^{load/PV})^{im}
\label{eq:q_element}
\end{equation}

Kirchoff's Current Law (KCL) for both real and imaginary part is ensured implementing (\ref{eq:kirchoff_current}).
\begin{equation}
    (I_{i,p,t }^{load})^{re/im} - (I_{i,p,t }^{PV})^{re/im} - I_{l,h \rightarrow i,p,t}^{re/im} + I_{l,i \rightarrow j,p,t}^{re/im} = 0
    \label{eq:kirchoff_current}
\end{equation}

In calculating an LV network's PV hosting capacity, it is important to define a set of technical constraints that will limit the generation of a PV system accordingly to values of phase voltage magnitudes, currents that flow through a transformer and LV lines, and voltage unbalance factor (VUF) in every node. Therefore, (\ref{eq:max_current})-(\ref{eq:volt_pos_square_mag}) are introduced. The maximum value of each line's current is defined as the input parameter and it changes depending on the type of a line or transformer. The minimum value of voltage magnitude is 0.9 p.u. and the maximum value is equal to 1.1 p.u. Based on definitions in the EN 50160 standard \cite{en50160} and the Croatian distribution grid code \cite{CroatianGridCode}, VUF cannot be larger than 2\%.

\begin{equation}
    (I_{l,ij,p,t}^{re})^2 + (I_{l,ij,p,t}^{im})^2 \leq (I_{l,ij}^{max})^2
    \label{eq:max_current}
\end{equation}

\begin{equation}
    (U^{min})^2 \leq (U_{n,p,t}^{re})^2 + (U_{n,p,t}^{im})^2 \leq (U^{max})^2
    \label{eq:node_volt_limit}
\end{equation}
\begin{equation}
    \frac{|U_{n,2,t}|^2}{|U_{n,1,t}|^2} \leq (VUF_n^{max})^2
    \label{eq:vuf}
\end{equation}

\begin{equation}
\begin{gathered}
    |U_{n,2,t}|^2=[U_{n,a,t}^{re}-\frac{1}{2}\cdot (U_{n,b,t}^{re} + U_{n,c,t}^{re})+ \\ \frac{\sqrt{3}}{2}\cdot (U_{n,b,t}^{im} - U_{n,c,t}^{im})]^2 +
    [U_{n,a,t}^{im}- \\ \frac{1}{2}\cdot (U_{n,b,t}^{im} + U_{n,c,t}^{im})-\frac{\sqrt{3}}{2}\cdot (U_{n,b,t}^{re} - U_{n,c,t}^{re})]^2
\end{gathered}
\label{eq:volt_neg_square_mag}
\end{equation}

\begin{equation}
\begin{gathered}
    |U_{n,1,t}|^2=[U_{n,a,t}^{re} -
    \frac{1}{2}\cdot (U_{n,b,t}^{re} + U_{n,c,t}^{re})- \\ \frac{\sqrt{3}}{2}\cdot (U_{n,b,t}^{im} - U_{n,c,t}^{im})]^2 + 
    [U_{n,a,t}^{im}-\\\frac{1}{2}\cdot (U_{n,b,t}^{im} + U_{n,c,t}^{im})+\frac{\sqrt{3}}{2}\cdot (U_{n,b,t}^{re} - U_{n,c,t}^{re})]^2
\end{gathered}
\label{eq:volt_pos_square_mag}
\end{equation}

\subsection{Determining a PV System Connection Phase}
PVs installed in LV networks can be single-phase or three-phase connected. Even though the single-phase connection is limited with relatively lower power, the implementation of such a connection is cheaper and simpler than a three-phase connection. The physical solution for ensuring a single-phase connection is easy to understand, however integration of the constraint in the mathematical model is not as intuitive and requires the introduction of binary variables. A single-phase connection of a PV system is implemented with \eqref{const:pv_power}-\eqref{const:pv_binaries}.

\begin{equation}
    P^{PV}_{n,p,t} \leq x_{n,p,t}^{PV} \cdot (P^{PV})^{max} \cdot PV_t^{curve}
    \label{const:pv_power}
\end{equation}

\begin{equation}
    \sum_{p \in \{a,b,c\}} x^{PV}_{n,p,t} \leq 1
    \label{const:pv_binaries}
\end{equation}

The maximum power $(P^{PV})^{max}$ of a single-phase connected PV system is equal to $3.68 kW$, as defined in the national distribution grid code \cite{CroatianGridCode}. However, the defined limit is a static one, meaning that its value remains the same, without considering network conditions such as network topology or demand. Moreover, the single-phase connection limit is relatively low, and larger deviations in terms of the value of the objective function are not expected. Therefore, we define a second scenario in which the maximum power $(P^{PV})^{max}$ is equal to $100 kW$, which is the maximum value of PV systems that can be connected to a Croatian LV network. Such an approach is also in line with the recent trend of abandoning static export limits and the use of dynamic operating envelopes instead \cite{YI2022108465}. Additionally, $PV_t^{curve}$ is the parameter created based on the irradiance and it further limits the PV generation but also ensures that there is no generation during the night or cloudy hours and that it is not possible to have maximum generation at all time periods.

Since the introduction of binary variables in the model uplifts the model from NLP to MINLP, the complexity of the model and an increase in the computational time are expected. To resolve the problem but still keep the NLP formulation, we propose four different approaches. 

In the first and the second approach, a PV system is connected to a randomly selected phase as shown with (\ref{eq:rand_phase})-(\ref{eq:pv_zero_rand}). The difference is that in one solution, the phase connection is randomly selected in every time period of the optimization process, and in the other, the connection phase is selected only once before the start of the optimization process and it remains the same in all time periods.

\begin{equation}
    q = random\{a,b,c\}
    \label{eq:rand_phase}
\end{equation}

\begin{equation}
    P^{PV}_{n,p=q,t} \leq (P^{PV})^{max} \cdot PV_t^{curve}
    \label{eq:pv_max_rand}
\end{equation}

\begin{equation}
    P^{PV}_{n,p \neq q,t} = 0
    \label{eq:pv_zero_rand}
\end{equation}

In the third approach, it is also assumed that the connection phase can change at every time period. However, unlike in previous solutions, the connection phase in this one is the same as the most loaded one, in order to maximally balance the total demand in every node. The final approach is similar to the previous one, with the exception of having the fixed PV system connection phase. In this approach, $D$ is determined as the maximum of sum of phase demand values over time. Eq. \eqref{eq:max_node_demand} presents determining the highest demand $D$ in the last two approaches. The method for determining the PV system connection phase $q$ is shown with Algorithm \ref{alg:pv_connect}. Constraining the maximum PV system generation power and ensuring single-phase connection is represented with \eqref{eq:pv_max_rand} and \eqref{eq:pv_zero_rand}.

\begin{equation}
\begin{gathered}
    D = \max \{P^{load}_{n,p=a,t}, P^{load}_{n,p=b,t}, P^{load}_{n,p=c,t}\}\\    
    D = \max \{\sum_{t \in T}P^{load}_{n,p=a,t}, \sum_{t \in T}P^{load}_{n,p=b,t}, \sum_{t \in T}P^{load}_{n,p=c,t}\}
    \label{eq:max_node_demand}
\end{gathered}
\end{equation}

\begin{algorithm}
\caption{Determining a PV system's variable connection phase}
\begin{algorithmic}
\If {$D = P^{load}_{n,p=a,t} \ or \  D = \sum_{t \in T}P^{load}_{n,p=a,t}$}
    \State {$q = a$}
\ElsIf{$D = P^{load}_{n,p=b,t} \ or \ D = \sum_{t \in T}P^{load}_{n,p=b,t}$}
    \State {$q = b$}
\ElsIf{{$D = P^{load}_{n,p=c,t} \ or \ D = \sum_{t \in T}P^{load}_{n,p=c,t}$}}
    \State {$q = c$}
\EndIf
\end{algorithmic}
\label{alg:pv_connect}
\end{algorithm}

Following the mathematical formulation of four different approaches, each of them is defined as an input in the presented OPF formulation used in the calculation of an LV network PV hosting capacity.

\subsection{Objective Function}
As previously stated, the aim of calculating the PV hosting capacity is to determine the maximum production of PVs, without violating any technical constraints in a distribution network. Therefore, the objective function is given in \eqref{obj:max_hc}. 

\begin{equation}
    \max \sum_{n \in N_{PV}}\sum_{p \in {a,b,c}}\sum_{t \in T} P_{n,p,t}^{PV} \cdot \frac{1}{4}h
    \label{obj:max_hc}
\end{equation}

Since demand measurements are collected in 15-minute intervals, the generation power of PV systems is calculated in the same time interval. Therefore, it is necessary to convert generation power $P_{n,p,t}^{PV}$ to energy using factor $\frac{1}{4}$, which is equal to 15 minutes.

Even though the hosting capacity is defined as the maximum amount of PVs or other technology that can be installed in a network without violating any technical constraints and endangering the safe and reliable operation of a network, we focused on maximizing the PV generation instead of installed power. Even though the objective functions of calculating PV hosting capacity and maximum PV generation are similar in their formulation, PV hosting capacity will not change as much as maximum generation due to the low PV power limit. Therefore, the objective function is as defined above.

Besides the values of the objective function in different case studies, we compare the computational time needed for solving the optimization problem, total active network losses, voltage magnitude, and VUF values.

\section{Case Study} \label{sec:cs}
\begin{figure}[htbp]
    \centering
    \includegraphics[width=\columnwidth]{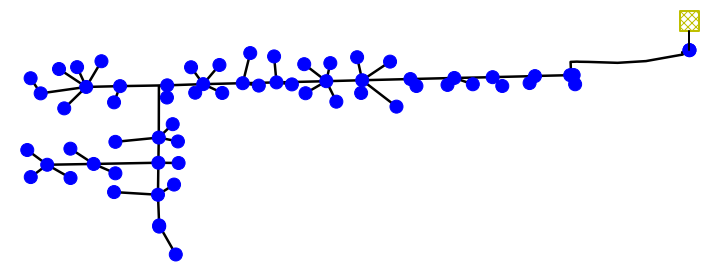}
    \caption{A residential LV feeder}
    \label{fig:net}
\end{figure}

The loss of accuracy caused by removing binary variables from the formulation used in calculating PV hosting capacity is tested on a mathematical model of a real-world residential LV feeder shown in  Fig. \ref{fig:net}. The feeder consists of one MV bus, 63 LV nodes, 62 underground cables, MV/LV transformer, and 64 unbalanced loads. Phase consumption curves for each end-user are created using measurements collected from smart meters installed in a feeder.

For the purpose of analyzing the impact of the PV hosting capacity formulation on the potential loss of accuracy, we define following case studies:
\begin{itemize}
    \item CS 1: MINLP formulation is used for determining the optimal connection phase of a PV system
    \item CS 2: connection phase of a PV system is randomly assigned in every time period
    \item CS 3: connection phase of a PV system is randomly assigned but is the same in every time period 
    \item CS 4: connection phase changes in each time period and it is equal to the most loaded phase
    \item CS 5: connection phase is the same in all time periods and it is equal with the highest summed value of active power
\end{itemize}

A formulation used in CS 1 is the only one ensuring the optimal solution of the hosting capacity calculation. However, as mentioned before, the use of binary variables is expected to make the problem even more complex. In order to tackle the complexity of the PV hosting capacity calculation, we introduce CS 2 - CS 5. Phase switching devices are utilized for balancing the network and removing unbalance problems in a network. Therefore, under the assumption that they are already installed, it is possible to consider the impact of the change of connection phase on the value of objective function and other investigated quantities, as presented in CS 1. The same assumption is introduced in CS 2 in which the use of binary variables is replaced by random selection of connection phase in every time period. That way, the OPF formulation remains nonlinear but the single-phase connection and the change of the connection phase remain ensured, which allows the comparison of the value of the objective function and other quantities. Since it is not realistic to expect that all end-users have installed switching devices that allow the change of connection phase, CS 3 in which the connection phase is randomly selected and remains the same during all time periods is introduced. In most cases end-users are not aware of the connection of their single-phase PV systems, therefore, CS 2 and CS 3 are the ones that most likely reflect a real-world situation. On the contrary, distribution system operators (DSOs) would like to prevent the random connection of single-phase PVs due to the possibility of an increase in network losses, voltage magnitude, or VUF values. For that reason, CS 4 and CS 5 are created. In CS 4, there is a possibility of changing the connection phase in different time periods, and a PV system is always connected to the phase with the highest demand. That way, the difference between the total demand of phases a, b, and c will be the lowest and the network will be the most balanced. The approach in CS 5 is more conservative and the change of demand phase is not possible. A PV system is connected to the phase with the highest daily demand. Even though this approach does not guarantee that the demand of a connection phase will be the highest in every time period, it is suitable for end-users that do not have installed phase-switching devices.

\section{Results} \label{sec:results}
Analyses in all case studies are done using Python 3.9.16 programming language and the pyomo optimization framework \cite{Hart2011}. The MINLP formulation in CS 1 is solved by the Knitro solver. Formulations in CS 2 - CS 5 are NLP without binary variables but knitro is also used in analyses. PC specifications are AMD Rzyen 5 3600 6-Core processor and 16.0 GB of RAM.

Table \ref{tab:results} summarizes the most important results of solving the optimization problem, values of the objective function and computational time for all case studies, and different constraints of the maximum PV power. When the maximum PV power is limited by the national grid code and is equal to $3.68 kW$, the best value of the objective function is in CS 1 since it is the only case study in which the exact optimal solution of the given hosting capacity problem is guaranteed. However, the computational time is higher than one hour, but despite the long duration, the calculation of maximal daily PV production is in between the planning and operation problems and the computational time does not present an obstacle in using the MINLP model. Values of the objective function in CS 2-CS 4 are between 15 kWh and 1 kWh smaller than in CS 1. The results clearly show that removing binary variable and using an NLP model cause the loss of accuracy, but on the other hand, the computational time needed for solving the optimization problems are less than 2 minutes, which significantly reduces the computation burden and computation cost. The results are not unambiguous and it is up to system operators to determine what is more important, to have an exact solution, or to use randomization and approximation approaches that will lead to a faster solving of this and other similar optimization problems.

Since the second analysis assumes the maximum connection power of single-phase PVs to be 100 kW, non-static export limits higher than 3.68 kW are expected, i.e., maximum PV power will vary at different nodes in a network. Therefore, the difference between the optimal solution in CS 1 and solutions in CS 2-CS 5 is expected to be much more significant. However, the given MINLP problem is too computationally complex, and the time needed to find the optimal solution is higher than 3 days. Since the computational time that long is not feasible and the approach is non-scalable, optimization was terminated after running for three days and there is no solution in CS 1. The value of the objective function in CS 2-CS 5 varies between 2 542 kWh and 2 608 kWh, which is not a large interval of values considering the maximum connection power constraint. Computational time in CS 2, CS 4, and CS 5 is around 2.5 minutes, which is significantly less than the computational time in CS 1. Even though the computational time in CS 3 is also less than in CS 1, it is more than three times higher when compared to CS 2, CS 4, and CS 5. This shows the importance and the impact of the connection phase on the optimal solution of the optimization problem and also shows the reason for the long duration of solving the MINLP formulation in CS 1.

\begin{table}[htbp]
\caption{Optimization results summary}
\resizebox{\columnwidth}{!}{%
\begin{tabular}{lccccc}
\cline{2-6}
\multicolumn{1}{l|}{}                                                                              & \multicolumn{5}{c|}{$(P^{PV})^{max} = 3.68 kW$}                                                                                                                          \\ \cline{2-6} 
\multicolumn{1}{l|}{}                                                                              & \multicolumn{1}{c|}{CS 1}                & \multicolumn{1}{c|}{CS 2}     & \multicolumn{1}{c|}{CS 3}     & \multicolumn{1}{c|}{CS 4}     & \multicolumn{1}{c|}{CS 5}     \\ \hline
\multicolumn{1}{|l|}{\begin{tabular}[c]{@{}l@{}}Objective \\ function \\ value (kWh)\end{tabular}} & \multicolumn{1}{c|}{893.87}              & \multicolumn{1}{c|}{878.72}   & \multicolumn{1}{c|}{892.60}   & \multicolumn{1}{c|}{882.22}   & \multicolumn{1}{c|}{889.41}   \\ \hline
\multicolumn{1}{|l|}{\begin{tabular}[c]{@{}l@{}}Computational\\  time (s)\end{tabular}}            & \multicolumn{1}{c|}{3 737.00}            & \multicolumn{1}{c|}{104.20}   & \multicolumn{1}{c|}{104.80}   & \multicolumn{1}{c|}{105.58}   & \multicolumn{1}{c|}{105.46}   \\ \hline
\multicolumn{1}{l|}{}                                                                              & \multicolumn{5}{c|}{$(P^{PV})^{max} = 100 kW$}                                                                                                                           \\ \cline{2-6} 
\multicolumn{1}{l|}{}                                                                              & \multicolumn{1}{c|}{CS 1}                & \multicolumn{1}{c|}{CS 2}     & \multicolumn{1}{c|}{CS 3}     & \multicolumn{1}{c|}{CS 4}     & \multicolumn{1}{c|}{CS 5}     \\ \hline
\multicolumn{1}{|l|}{\begin{tabular}[c]{@{}l@{}}Objective \\ function \\ value (kWh)\end{tabular}} & \multicolumn{1}{c|}{\textbf{---}}                   & \multicolumn{1}{c|}{2 607.58} & \multicolumn{1}{c|}{2 542.79} & \multicolumn{1}{c|}{2 608.35} & \multicolumn{1}{c|}{2 543.85} \\ \hline
\multicolumn{1}{|l|}{\begin{tabular}[c]{@{}l@{}}Computational\\  time (s)\end{tabular}}            & \multicolumn{1}{c|}{\textgreater 3 days} & \multicolumn{1}{c|}{149.86}   & \multicolumn{1}{c|}{464.01}   & \multicolumn{1}{c|}{144.60}   & \multicolumn{1}{c|}{135.75}   \\ \hline
                                                                                                   & \multicolumn{1}{l}{}                     &                               &                               &                               &                              
\end{tabular}}
\label{tab:results}
\end{table}

A more detailed analysis of optimization results includes analyses of active total daily network losses, voltage magnitude, and voltage unbalance. However, these analyses are conducted only for the first case in which the connection PV power is limited to 3.68 kW since the case of the increased connection power did not lead to the optimal solution within a reasonable time.

\begin{figure}[htbp]
    \centering
    \includegraphics[width=\columnwidth]{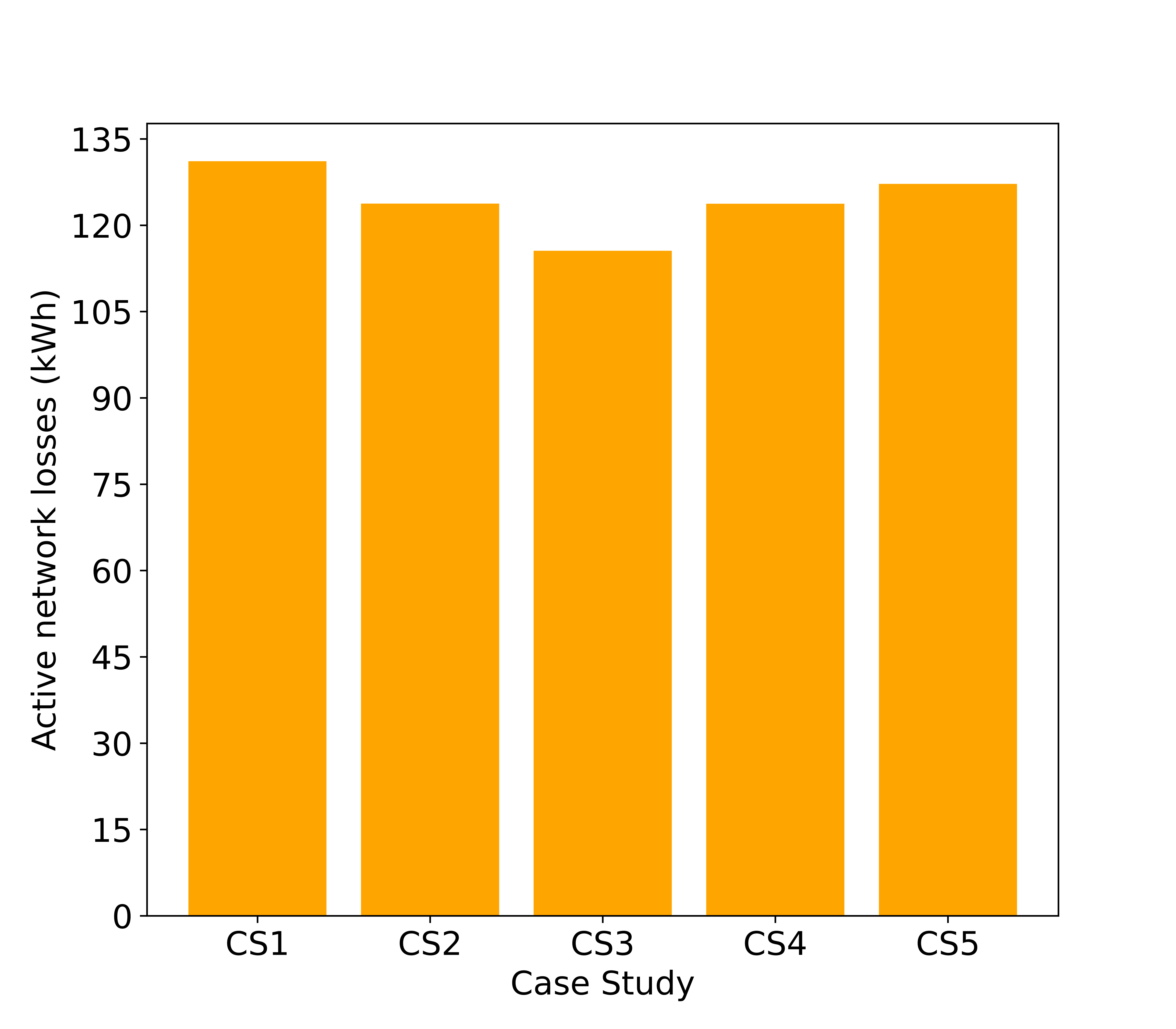}
    \caption{Total daily active network losses - CS 1}
    \label{fig:losses}
\end{figure}

Fig. \ref{fig:losses} shows the value of total daily network losses in CS 1-CS 5. As can be seen in the graph, the largest losses of 131.13 kWh occur in CS 1, when the daily production of PVs is calculated by using the MINLP formulation. Even though this solution leads to the maximization of production, it also increases the current that flows through lines due to the occurrence of reverse power flows. Losses in CS 2-CS 5 are in the range of 115.56-127.20 kWh and since the connection phase changes over case studies, the results show the correlation between the connection phase of a PV system and network losses. However, more general conclusions cannot be made without introducing stochasticity that allows defining multiple scenarios with different layouts of connecting PV systems.

\begin{figure}[htbp]
    \centering
    \includegraphics[width=\columnwidth]{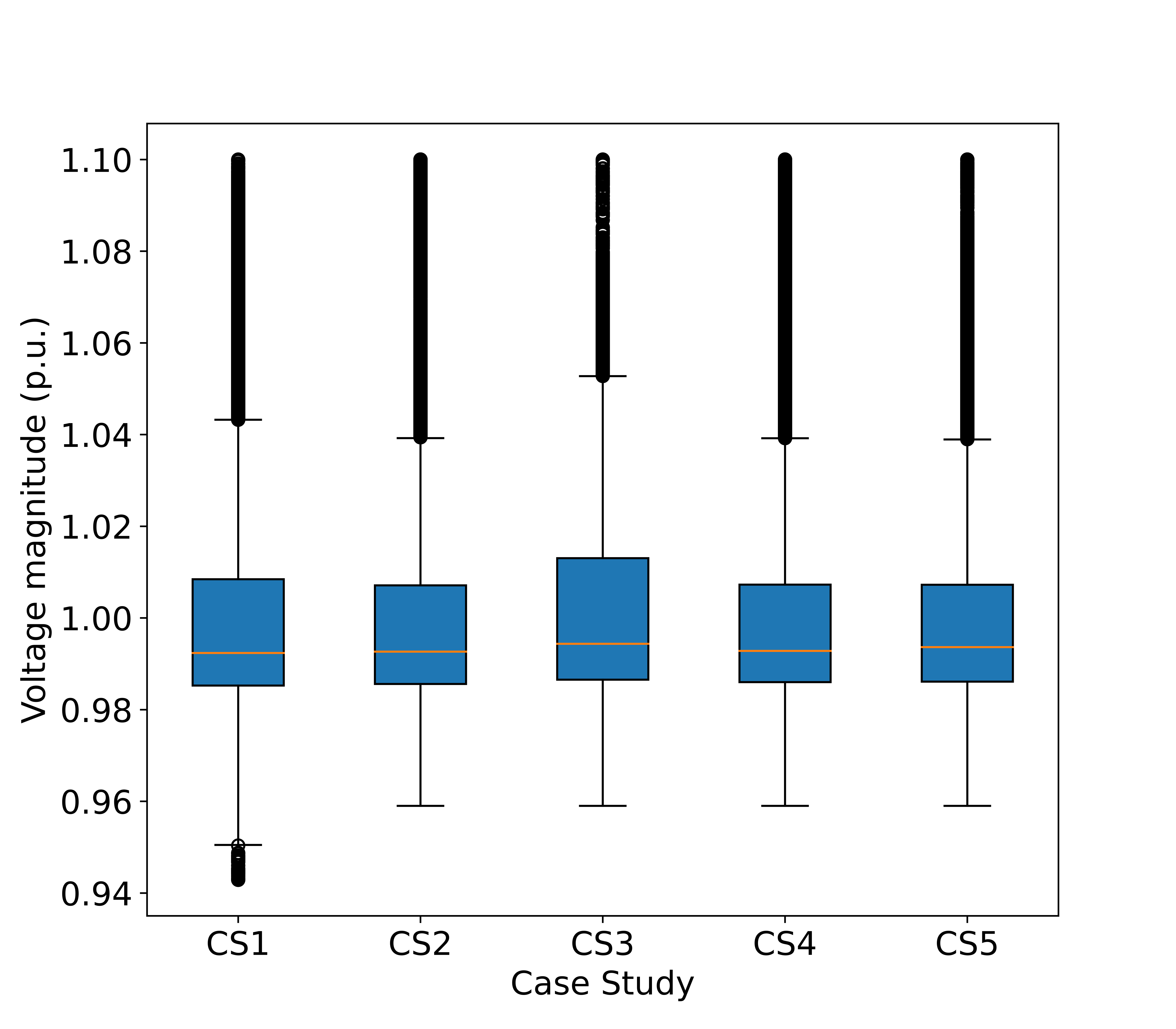}
    \caption{Voltage magnitude - CS 1}
    \label{fig:voltage}
\end{figure}

Boxplots representing voltage magnitude are shown in Fig. \ref{fig:voltage}. The upper bound of voltage magnitude is set to be 110\% of the nominal voltage, and as it can be seen, this value is reached in all case studies, even though the values that high are identified as outlier values and occur rarely in comparison to the value in the interval 0.95 p.u.-1.05 p.u. The median value is around 0.99 p.u. and is almost the same in all case studies. Without considering outlier values, the minimum voltage magnitude in CS 1 is around 0.95 p.u. and around 0.96 p.u. in CS 2-CS 5. Maximum voltage values are in the range of 1.04 p.u.-1.06 p.u. The results show that even though the connection phase impacts voltage magnitude, values in the graph do not significantly vary in different case studies. 

\begin{figure}[htbp]
    \centering
    \includegraphics[width=\columnwidth]{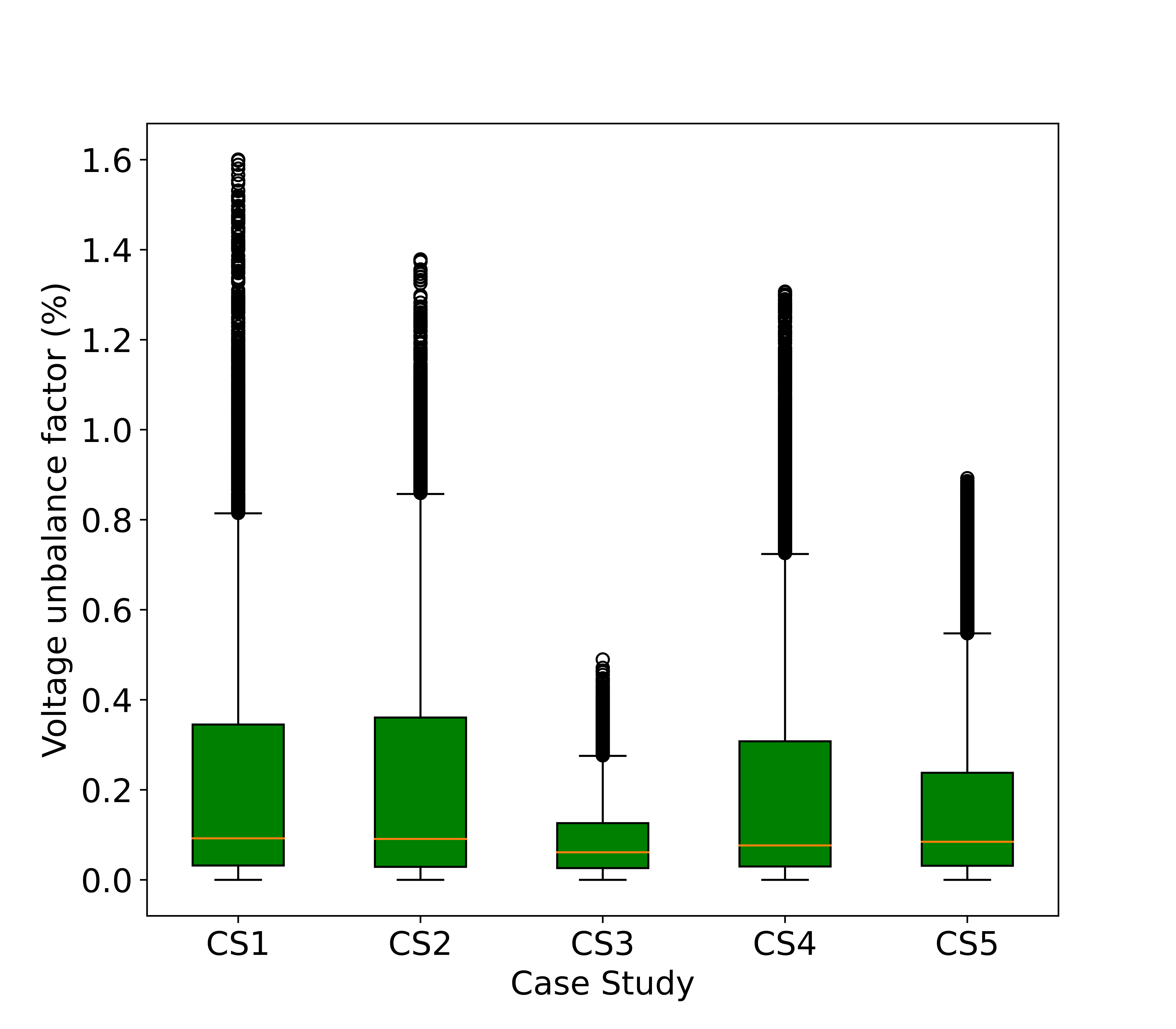}
    \caption{Voltage unbalance factor - CS 1}
    \label{fig:vuf}
\end{figure}

Values of VUF are presented in Fig. \ref{fig:vuf}. As it can be seen, values in the graph do not go higher than 1.6\% and since the threshold VUF value is 2\%, PV systems constrained by the connection power of 3.68 kW are not additionally constrained by voltage unbalance. Another interesting fact is the correlation of VUF with the value of network losses in CS 1-CS  5. When considering all values in a boxplot together with outlier values, the highest VUF occurs in CS 1 which corresponds to the value of network losses in the same case study. On the contrary, the lowest values of VUF and network losses are in CS 3. From these results, it is possible to determine a direct impact of voltage unbalance on network losses and to conclude that losses can be reduced by minimizing unbalance in a network.

\section{Conclusion} \label{sec:conclusion}
The share of installed PV systems in LV networks is continuously increasing. There are numerous existing methods that focus on determining the PV hosting capacity and many of them are based on linearization or approximation optimization approaches that often lead to a sub-optimal solution. To overcome the problem, we use a non-linear OPF model in calculating the maximum daily production of single-phase PV systems. Additionally, we uplift the problem to the MINLP formulation, the only formulation that guarantees an exact, optimal solution to the given optimization problem. The MINLP formulation is used in two different cases, the first approach in which the PV system is constrained by the national grid code (3.68 kW) and the second approach in which connection power is constrained with a non-static limit. The computational time needed to solve the problem is higher than one hour in the first case, and in the second case optimization was terminated after it was not able to find the optimal solution for three days. To decrease the computational burden and decrease the duration of optimization, four different methods for determining the connection phase that do not use binary variables were used. The results show an expected decrease in the computational time and the loss of accuracy in terms of the objective function value. Moreover, a more detailed analysis of network losses, voltage magnitude, and voltage unbalance factor was conducted to further investigate the difference when using different approaches. Analyses show that the range of values of different quantities varies based on the observed case study, but differences are not significant and cannot be determined as obstacles in using one of the proposed NLP approaches.

Future work will focus on investigating scalability problems in optimization problems in three-phase LV networks. Problems include the introduction of stochasticity in hosting capacity problems, the use of linearized or approximated OPF models, but also considering other distributed energy resources in the optimization problem..

\bibliographystyle{IEEEtran}
\bibliography{IEEEabrv,references}

\end{document}